\documentclass[%
 reprint,
 amsmath,amssymb,
 aps,
]{revtex4-2}

\usepackage{graphicx}% Include figure files
\usepackage{dcolumn}% Align table columns on decimal point
\usepackage{bm}% bold math
\begin{document}

\preprint{APS/123-QED}

\title{Parameter Dependence and Bell nonlocality}% Force line breaks with \\

\author{Moji Ghadimi}
 \email{moji131@gmail.com}

\affiliation{%
Center for Quantum Dynamics, Griffith University, Nathan QLD Australia 
}%

% \begin{abstract}
% Bell's theorem asserts that no model that satisfies all of the plausible physical assumptions of outcome independence (OI), measurement independence (MI) and Parameter Independence (PI) can reproduce quantum mechanics. Here I find the optimum model that saturates CHSH inequality for the case that outcome independence and measurement independence hold but parameter dependence is allowed. The symbolic optimizations to find the saturating models are performed using Analytical Optimizer v1.0. 
% \end{abstract}

\begin{abstract}
Bell's theorem asserts that no model that satisfies all of the plausible physical assumptions of outcome independence (OI), measurement independence (MI) and Parameter Independence (PI) can reproduce quantum mechanics. Here I find the optimum model that saturates CHSH inequality for the case that outcome independence and measurement independence hold but parameter dependence is allowed. I find the bound as a function of a measure of parameter dependence and show that the model proposed by [New Journal of Physics,12(8):083051] is optimal for one-way parameter dependence. The symbolic optimizations to find the saturating models are performed using the open source software Analytical Optimizer v1.0. 
\end{abstract}
\maketitle

%\tableofcontents

\section{Introduction}
Bell inequalities demonstrate that any model of some quantum mechanical experiments performed in space-like separated regions cannot satisfy all of the physically plausible properties of outcome independence (OI), measurement independence (MI) and Parameter Independence (PI) at the same time \cite{bell_einstein_1964, clauser_proposed_1969}. The violation of these inequalities by certain quantum correlations, implies that at least one such property must be relaxed by any model of these correlations. Finding the minimum amount of relaxation needed to achieve some violation of these inequalities is helpful for construction of fundamental theories of nature and also for quantifying resources in quantum computing and communication. 

Minimum relaxation requirement of some of these assumptions to simulate quantum mechanics has been investigated in the literature \cite{toner_communication_2003, branciard_testing_2008, barrett_how_2011, hall_local_2010, brask_bell_2017,hall_relaxed_2011}. For example, it is known that to simulate a singlet state 1 bit of randomness generation or outcome dependence, 1 bit of signaling or communication (parameter dependence), or 1/15 of one bit of correlation between the underlying variable and the measurement settings (measurement dependence) is needed \cite{hall_relaxed_2011}.

In this paper I investigate the upper bound for violation of the Clauser-Horne-Shimony-Holt (CHSH) inequality for any given relaxation of parameter independence. I write the right hand side of the CHSH inequality as a function of joint outcome probabilities and then optimise it under constraint of the measure of parameter dependence corresponding to the maximum possible change in a probability in one region due to a measurement made in a distant region. To find the optimum value of this function subject to the constraints mentioned, I use Analytical Optimizer v1.0 (see Appendix \ref{ap_aoc}).   

I find a relationship between the sturating bound of the CHSH inequality and a measure of parameter dependence for the case of one-way and two-way dependence.
The result answers the question posed in \cite{hall_relaxed_2011} and proves that the one way communication model of Pawlowski et al.  \cite{pawlowski_non-local_2010,hall_relaxed_2011} is optimal.

\section{A Measure of Parameter Dependence}

The CHSH scenario considers an experiment with two settings and two outcomes for measurements performed by distant experimenters, Alice and Bob say, in space-like separated regions. For some fixed preparation procedure, a set of statistical correlations, ${p(a, b|x, y)}$, can be assigned to this experiment where the pair $(a, b)$ labels the possible outcomes and the pair $(x, y)$ labels possible experiment settings. Any underlying model of these correlations introduces an underlying variable $\lambda$ on which the correlations depend, which is typically interpreted as representing information about the preparation procedure. From Bayes theorem one has the identity
\begin{align}
p(a, b|x, y) = \int d\lambda p(a, b|x, y, \lambda) p(\lambda|x, y),
\label{eq_prob}
\end{align}
with integration replaced by summation over any discrete ranges of $\lambda$. A given underlying model specifies the type of information encoded by $\lambda$, and the underlying probability densities $p(a, b|x, y, \lambda)$ and $p(\lambda|x, y)$.
Definitions and quantitative measures of relaxation for each of parameter independence, outcome independence and measurement independence are introduced in \cite{hall_relaxed_2011}. Based on those definitions MI can be simply defined as  $p(\lambda|x, y) = p(\lambda)$ that is probabilities of preparations are independent of measurement settings. OI maybe defined as the property that measurement outcomes are uncorrelated given the knowledge of the underlying variable. Outcome dependence then may be defined as maximum variational distance between an underlying joint distribution and the product of its  marginals \cite{hall_relaxed_2011}
\begin{align}\begin{split}
OD :=
\underset{x,y,\lambda}{\sup} 
 \sum_{a, b} \lvert p(a, b|x, y, \lambda) - p(a|x, y, \lambda)\\ p(b|x, y, \lambda) \rvert 
 \label{eq_od_def}
\end{split}\end{align}
where $0 \leq OD \leq 2$ and $OD=0$ if and only if outcome independence is satisfied. Here I do not relax outcome independence assumption therefore I only use $OD=0$ as a constraint. 

The property that we intend to relax is parameter independence. Parameter independence holds when the underlying marginal distribution associated with one outcome is independent of the other parameter. A measure of parameter dependence can be maximum possible change in an underlying marginal probability for one party, as the consequence of changing the measurement parameter of the other party. Following \cite{hall_relaxed_2011} and \cite{ hall_complementary_2010} we can define parameter dependence ($\rm PD$) from Alice to Bob as
\begin{align}
{\rm PD}_{A \rightarrow B} := \underset{x,x',y,b, \lambda}{\sup} |p(b|x, y, \lambda) - p(b|x', y, \lambda)|,
\end{align}
similarly for the other direction we have
\begin{align}
{\rm PD}_{B \rightarrow A} := \underset{x,y,y',b, \lambda}{\sup} |p(a|x, y, \lambda) - p(a|x, y', \lambda)|. 
\end{align}
The two way parameter dependence allowed can be defined as
\begin{align}
{\rm PD} := \max \{ {\rm PD}_{A \rightarrow B}, {\rm PD}_{B \rightarrow A} \}.
\label{sig_2w}
\end{align}
Here $0 \leq {\rm PD} \leq 1$ and ${\rm PD}=0$ if and only if parameter independence is satisfied. This is a measure of maximum possible change in an underlying marginal probability of one experimenter, as the result of changing the measurement setting of the other observer.
A CHSH type inequality for an underlying model having value of parameter dependence PD can be written as
\begin{align}
\langle XY \rangle + \langle XY' \rangle + \langle X'Y \rangle - \langle X'Y' \rangle \leq B( {\rm PD})
\label{equ_chsh}
\end{align}
where $\langle XY \rangle$ denotes the average product of the measurement outcomes for setting $X$ for Alice and $Y$ for Bob. It is well-known that $B( {\rm PD}=0)$ (assuming also measurement independence and outcome independence) is 2 at maximum and B can reach $2\sqrt{2}$ in certain quantum mechanical experiments.

\section{Relaxed CHSH Inequalities for One-way and Two-way Parameter Dependence}
Here I find the maximum possible violation of CHSH inequality assuming outcome independence ($\rm OD = 0$ in Equ. \ref{eq_od_def}) and measurement independence ($\rm MI$) but allowing parameter dependence $(0\leq PD \leq 1)$. $\rm MI$ corresponds to $p(\lambda|x,y)=p(\lambda)$ in Equ.\ref{eq_prob}. I consider both one-way and two-way parameter dependence.

For the case of two-valued measurements, I denote the possible outcomes by $a,b = \pm 1$ and possible measurements by $x$ or $x'$ and $y$ or $y'$. If we define $c = p(a=+1,b=+1)$, $m = p(a=+1)$ and $n = p(b=+1)$,  for joint measurement setting $(x, y)$, the corresponding joint measurement distribution can be written in the form
\begin{align}
\begin{split}
p_1(a=+1, b=+1|x, y, \lambda) =\\ c_1(\lambda) \\
p_1(a=+1, b=-1|x, y, \lambda) =\\ m_1(\lambda) - c_1(\lambda)\\
p_1(a=-1, b=+1|x, y, \lambda) =\\ n_1(\lambda) - c_1(\lambda)\\
p_1(a=-1, b=-1|x, y, \lambda) =\\ 1 + c_1(\lambda) - m_1(\lambda) - n_1(\lambda).
\label{equ_joint_p}
\end{split}
\end{align}
Similarly we use subscript 2 to define $c_2(\lambda)$, $m_2(\lambda)$ and $n_2(\lambda)$ for joint  measurement setting $(x, y')$, 3 for $(x', y)$ and 4 for $(x', y')$.

If we define left hand side of equation \ref{equ_chsh} to be
\begin{align}
E_{\lambda} = \langle XY \rangle_\lambda + \langle XY' \rangle_\lambda + \langle X'Y \rangle_\lambda - \langle X'Y' \rangle_\lambda  
\label{equ_ex_2}
\end{align}
using Equ. \ref{equ_joint_p}, for any given $\lambda$, we have
\begin{align}
\langle XY \rangle_\lambda  =  1 + 4c_1 - 2m_1 - 2n_1 .
\label{equ_ex_xy}
\end{align}
And similarly we can use labels 2, 3 and 4 for $\langle XY' \rangle_\lambda$, $\langle X'Y \rangle_\lambda$ and $\langle X'Y' \rangle_\lambda$ respectively. Using Equ. \ref{equ_ex_xy}, Equ. \ref{equ_ex_2} becomes
\begin{align}
\begin{split}
E_{\lambda}  =  (1 + 4c_1 - 2m_1 - 2n_1)\\ + (1 + 4c_2 - 2m_2 - 2n_2)\\ +
 (1 + 4c_3 - 2m_3 - 2n_3)\\ - (1 + 4c_4 - 2m_4 - 2n_4).
\end{split}
\label{ex_MD}
\end{align}
To find the upper bound, we need to maximize $E_{\lambda}$ with three constraints below:
\begin{enumerate}
 \item Since all the probabilities in Equ. \ref{equ_joint_p} needs to be between 0 and 1 and the sum of those needs to be 1 we have \begin{align}
\begin{split}
0 \leq m_i \leq 1\\
0 \leq n_i \leq 1\\
\max\{0,m_i+n_i-1\} \leq c_i \leq \min\{m_i,n_i\}\\
i= 1, 2, 3, 4.
\label{equ_pos}
\end{split}
\end{align}

 \item Outcome independence constraint:
\begin{align}
\begin{split}
|c_i-m_in_i| = 0 \\
i= 1, 2, 3, 4.
\label{eq_OD}
\end{split}
\end{align}
With this $c_i$ can be eliminated and the third constraint above is automatically satisfied.

 \item Parameter dependence constraint:
 This can either be the one-way parameter dependence constraint (allowing parameter dependence $2 \rightarrow 1$)
\begin{align}
\begin{split}
|m_1-m_2|,|m_3-m_4| \leq {\rm PD}\\
|n_1-n_3|,|n_2-n_4| = 0
\label{equ_s_1w}
\end{split}
\end{align}
or the two-way parameter dependence constraint
\begin{align}
\begin{split}
|m_1-m_2|,|m_3-m_4| \leq {\rm PD}\\
|n_1-n_3|,|n_2-n_4| \leq {\rm PD}
\label{equ_s_2w}
\end{split}
\end{align}

\end{enumerate}
The symbolic optimization of this equation was done using Analytical Optimizer v1.0 (see Appendix \ref{ap_aoc}). The results are summarised in the following sections.

\begin{table}
\begin{center}
\begin{tabular}{ | c |  c c c c |} 
 \hline
 Outcomes & $+,+$ & $+,-$ & $-,+$  & $-,-$\\
   \hline 
 Measurements &  &  &   & \\
\cline{1-1}
$x,y$ & 1 & 0 & 0  & 0\\
$x,y'$ & 0 & $1-\rm PD$ & 0  & $\rm PD$\\
$x',y$ & 1 & 0 & 0  & 0\\
$x',y'$ & 0 & 1 & 0  & 0\\
 
 \hline
\end{tabular}
\label{tab:ow} 
\caption{A saturating model for one-way parameter dependence. $x$ and $x'$ are measurement choices for one observer and $+$ and $-$ signs on the left side of each pair shows that observer's outcome. The scond observer has choices $y$ and $y'$ and signs on the right of the pair of signs, show this observers outcome.}
\end{center}
\end{table}

\subsection{One-way Parameter Dependence }
\begin{figure}
    \includegraphics[width=0.5\textwidth]{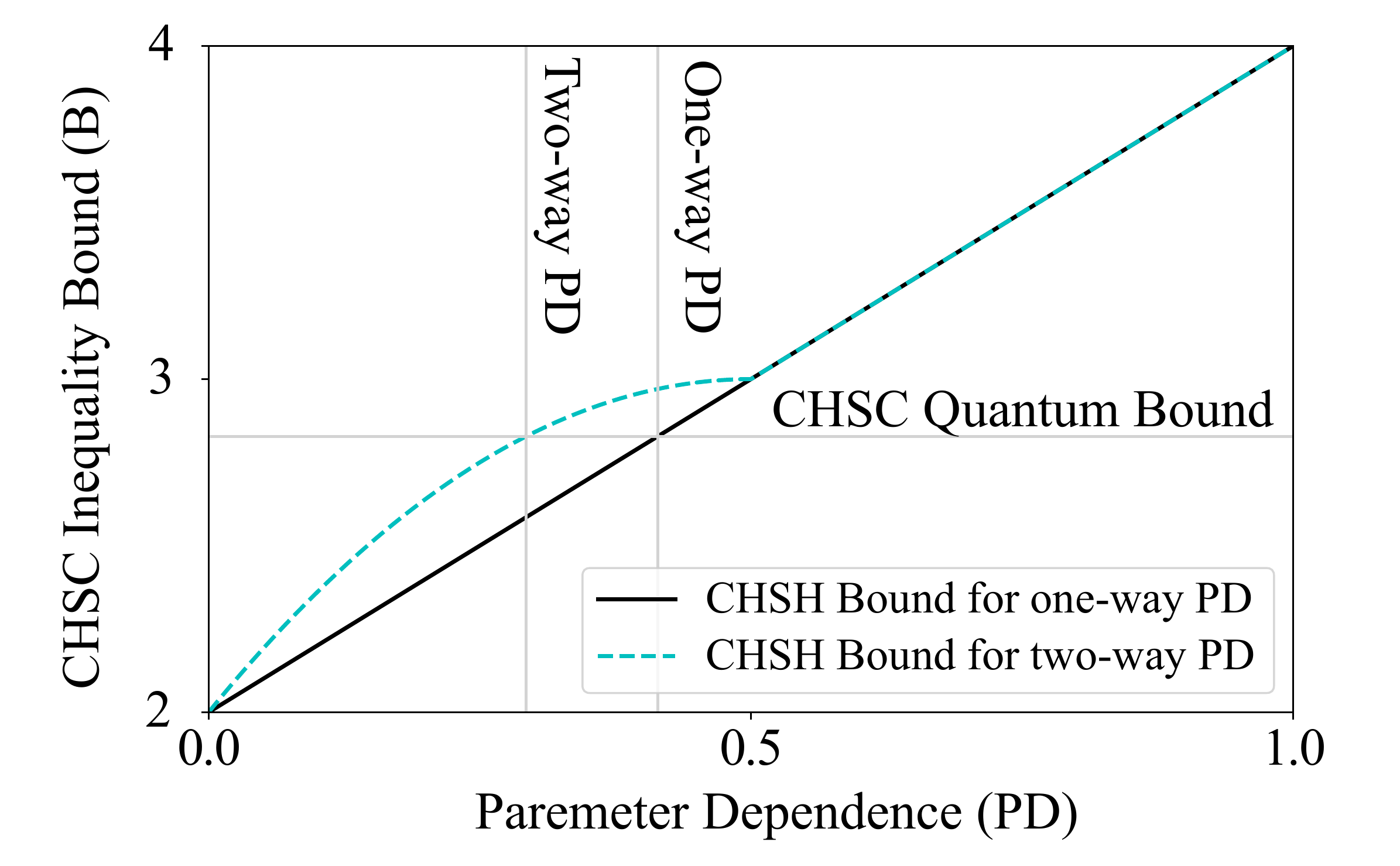}
    \caption{The bounds for CHSH inequality for different values of one-way and two-way parameter dependence (PD). The horizontal line shows the quantum (Tsirelson) bound for the CHSH inequality. Vertical lines show the value of $\rm PD$ required to simulate quantum mechanics for one-way and two-way parameter dependence. }
    \label{fig:1}
\end{figure}

For the case of one-way parameter dependence  the maximum value that can be found for $B({\rm PD})$ is (see Appendix \ref{ap_1ws})
\begin{align}
B_{ow}({\rm PD}) = 2{\rm PD}+2  \;\; for \;\; 0 \leq {\rm PD} \leq 1.
\end{align}
This means that to simulate quantum mechanics one-way parameter dependence  of at least ${\rm PD} = \sqrt{2}-1 \approx 0.414$ is required (see Fig. \ref{fig:1}).

One possible model to saturate this bound is shown in table 1 (see Appendix \ref{ap_1ws}). Note that these values are for one particular $\lambda$ and if we only have this one $\lambda$, the model allows signalling at the observer level. It is trivial to prevent signalling by taking equal mixture of this model and one with the outcomes flipped \cite{hall_measurement-dependence_2020}. This does not change the value of ${\rm PD}$. This argument applies to all the models in this paper.   
The result proves that the one way communication model of Pawlowski et al.  \cite{pawlowski_non-local_2010,hall_relaxed_2011}
\begin{align}
\begin{split}
 p(a, b|x, y, \lambda) =  p(a, b|x, y', \lambda) =  p(a, b|x', y, \lambda)\\ = \delta_{a\lambda} \delta_{b\lambda}\\
 p(a, b|x', y', \lambda) = [p(1-\delta_{a\lambda})+ (1-p)\delta_{a\lambda}]\delta_{b\lambda}, 
\end{split}
\end{align}
is optimal. Here $\lambda=\pm1$ and $p(\lambda) = \frac{1}{2}$, and it is straightforward to calculate that for arbitrary $p\in [0,1]$, PD = p and the violation of CHSH inequality is 2p (the same as the optimum found here).

\begin{table}
\begin{center}
\begin{tabular}{ | c |  c c |} 
 \hline
 Outcomes & $+,+$ & $+,-$ \\
   \hline 
 Measurement &  &  \\
\cline{1-1}  
$x,y$ & 1 & 0\\
$x,y'$ & \rm PD(1-\rm PD) & $\rm PD^2-2\rm PD+1$  \\
$x',y$ & 1 & 0 \\
$x',y'$ & 0 & 1 \\

  \hline
  
  \hline
 Outcomes &  $-,+$  & $-,-$\\
   \hline 
 Measurements   &   & \\
\cline{1-1}  
$x,y$ &  0  & 0\\
$x,y'$  & $\rm PD^2$ & $\rm PD-\rm PD^2$ \\
$x',y$ & 0 & 0\\
$x',y'$ & 0 & 0\\

  \hline
\end{tabular}
\label{tab:tw1} 
\caption{A saturating model for two-way parameter dependence, $0 \leq \rm PD \leq \frac{1}{2}$. $x$,$x'$,$y$ and $y'$ are measurement choices and $+$ and $-$ are outcomes. The left side of each pair corresponds to one observer and the right side to the other. }

\end{center}
\end{table}

\subsection{Two-way Parameter Dependence }
For the case of two-way parameter dependence  the optimum value for B is (see Appendix \ref{ap_2ws}) 
\begin{align}
B_{tw}({\rm PD}) = 4{\rm PD}(1 - {\rm PD}) + 2  \;\; for \;\; 0 \leq {\rm PD} \leq \frac{1}{2} \\
= 2{\rm PD} + 2 \;\; for \;\; \frac{1}{2} < {\rm PD} \leq 1.
\end{align}
This shows that if two-way parameter dependence  is allowed there will be an advantage for the case $0 \leq
{\rm PD} \leq \frac{1}{2}$ comparing to one-way parameter dependence  but for $\frac{1}{2} \leq  {\rm PD}\leq 1$ no advantage can be gained (See Fig. \ref{fig:1}). 
To simulate quantum mechanics ${\rm PD} = (1-\sqrt{3-2\sqrt{2}} ) /2 \approx 0.2929$ is required that is lower than what is needed for one-way parameter dependence (see Fig. \ref{fig:1}).

\begin{table}
\begin{center}
\begin{tabular}{ | c |  c c c c |} 
 \hline
 Outcomes & $+,+$ & $+,-$ & $-,+$  & $-,-$\\
   \hline 
 Measurements &  &  &   & \\
\cline{1-1} 
$x,y$ & \rm PD & 1-\rm PD & 0  & 0\\
$x,y'$  & 1 & 0 & 0  & 0\\
$x',y$ & 0 & 0 & 0 & 1\\
$x',y'$ & 0 & 0 & 1 & 0\\
 
  \hline

\end{tabular}
\label{tab:tw2} 
\caption{A saturating model (outcome probabilities for different measurement settings) for two-way parameter dependence, $\frac{1}{2} \leq \rm PD \leq 1$. }

\end{center}
\end{table}

Two possible saturating models for two-way parameter dependence for the cases of $0 \leq \rm PD \leq \frac{1}{2}$ and $\frac{1}{2} \leq \rm PD \leq 1$ are shown in tables 2 and 3 respectively.

\section{Conclusion}
Here I found the upper bound for a CHSH inequality when outcome independence and measurement independence are satisfied but parameter independence is not. I found the saturating bound for both one-way and two-way parameter dependence  and compared the bound with previously published models of communication. For future work it is of interest to find $B(\rm OD, \rm PD=0, \rm MD=0)$ and $B(\rm OD, \rm PD, M=0)$. One problem with $B(\rm OD, \rm PD=0, \rm MD=0)$ is that outcome dependence (Eq. \ref{eq_od_def}) is a quadratic constraint and the current version of Analytical Optimizer is not designed to handle quadratic constraints. An additional problem with $B(\rm OD, \rm PD, \rm MD=0)$ is that the current version of the code does not support more than one constant in the constraints but here we have both $\rm OD$ and $\rm PD$ as free parameters. 

\section*{Acknowledgement}
I thank Michael J. W. Hall, Mirko Lobino and Tim Gould for several helpful discussions.

\bibliographystyle{unsrt}

\bibliography{main.bib}% Produces the bibliography via BibTeX.

\appendix

\section{One-way Parameter Dependence Bound}
\label{ap_1ws}
For the case of one-way parameter dependence  using Equ. \ref{eq_OD} and equlities in Equ. \ref{equ_s_1w}, we can eliminate $c_i$, $n_3$ and $n_4$. Therefore Equ. \ref{ex_MD} becomes
\begin{align}
\begin{split}
E_{\lambda}  =  (1 + 4m_1n_1 - 2m_1 - 2n_1)\\ + (1 + 4m_2n_2 - 2m_2 - 2n_2)\\ + (1 + 4m_3n_1 - 2m_3 - 2n_1)\\ - (1 + 4m_4n_2 - 2m_4 - 2n_2) .
\end{split}
\label{equ_e_MD_1w}
\end{align}
Expanding rest of the constraints we have
\begin{align}
\begin{split}
m_1-m_2 \leq {\rm PD}, 
m_2-m_1 \leq {\rm PD},\\ 
m_3-m_4 \leq {\rm PD}, 
m_4-m_3 \leq {\rm PD}\\
0 \leq m_1 \leq 1, 
0 \leq m_2 \leq 1, 
0 \leq m_3 \leq 1\\
0 \leq m_4 \leq 1,
0 \leq n_1 \leq 1,
0 \leq n_2 \leq 1.
\end{split}
\end{align}

To find the maximum value of $E_{\lambda}$ in Equ. \ref{equ_e_MD_1w} subject to the constraints above I used the code Analytical Optimizer (see Appendix \ref{ap_aoc}). This code finds the optimum to be and $2\rm PD + 2$ for $0 \leq \rm PD \leq 1$. 

Values for one possible set of input parameter to achieve the presented maximum function value is listed in table 4 which leads to the model shown in table 1.

\begin{table}
\begin{center}
\begin{tabular}{ | c |  c c c  |} 
\hline
   & c & m & n \\
  \hline  
1 & 1 & 1 & 1  \\
2  & 0 & 1-\rm PD & 0  \\
3 & 1 & 1 & 1 \\
4 & 0 & 1 & 0 \\
\hline 

\end{tabular}
\label{tab:ow_opt} 
\caption{One possible set of optimal input values for one-way parameter dependence.}

\end{center}
\end{table}

\section{Two-way Parameter Dependence Bound}
\label{ap_2ws}
For the case of two-way parameter dependence  using Equ. \ref{eq_OD} we can eliminate $c_i$ and Equ. \ref{ex_MD} becomes
\begin{align}
\begin{split}
E_{\lambda}  =  (1 + 4m_1n_1 - 2m_1 - 2n_1)\\ + (1 + 4m_2n_2 - 2m_2 - 2n_2)\\ +
 (1 + 4m_3n_3 - 2m_3 - 2n_3)\\ - (1 + 4m_4n_4 - 2m_4 - 2n_4) .
\end{split}
\label{equ_e_MD}
\end{align}
Expanding rest of the constraints we have
\begin{align}
\begin{split}
m_1-m_2 \leq {\rm PD}, 
m_2-m_1 \leq {\rm PD}, 
m_3-m_4 \leq {\rm PD},\\ 
m_4-m_3 \leq {\rm PD},
n_1-n_3 \leq {\rm PD}, 
n_3-n_1 \leq {\rm PD},\\ 
n_2-n_4 \leq {\rm PD}, 
n_4-n_2 \leq {\rm PD},
0 \leq m_1 \leq 1,\\ 
0 \leq m_2 \leq 1, 
0 \leq m_3 \leq 1, 
0 \leq m_4 \leq 1,
0 \leq n_1 \leq 1,\\ 
0 \leq n_2 \leq 1, 
0 \leq n_3 \leq 1, 
0 \leq n_4 \leq 1
\end{split}
\end{align}
Analytical Optimizer (Appendix \ref{ap_aoc}) finds the maximum value of $E_{\lambda}$ in Equ. \ref{equ_e_MD} to be $4\rm PD(1 - \rm PD) + 2$ for $0 \leq \rm PD \leq \frac{1}{2}$ and $2\rm PD + 2$ for $\frac{1}{2} \leq \rm PD \leq 1$.

For $0 \leq \rm PD \leq \frac{1}{2}$ one possible saturating model can be achieved with values in table 5. The corresponding model is shown in table 2. For $\frac{1}{2} \leq \rm PD \leq 1$ a saturating model can be made with values in table 6 and the resulting model is shown in table 3.

\begin{table}
\begin{center}
\begin{tabular}{ | c |  c c c  |} 
\hline
   & c & m & n \\
  \hline  
1 & 1 & 1 & 1  \\
2  & \rm PD(1-\rm PD) & (1-\rm PD)  & \rm PD \\
3 & 1 & 1 & 1 \\
4 & 0 & 1 & 0 \\
\hline 

\end{tabular}
\label{tab:tw_opt_1} 
\caption{A set of input values that maximizes the bound function for the two-way parameter dependence for the case of $0 \leq \rm PD \leq \frac{1}{2}$.}

\end{center}
\end{table}

\begin{table}
\begin{center}
\begin{tabular}{ | c |  c c c  |} 
\hline
   & c & m & n \\
  \hline  
1 & \rm PD & 1 & \rm PD  \\
2 & 1 & 1 & 1 \\
3 & 0 & 0 & 0 \\
4 & 0 & 0 & 1 \\
\hline 

\end{tabular}
\label{tab:tw_opt_2} 
\caption{One example of optimal input values for parameter dependence when $\frac{1}{2} \leq \rm PD \leq 1$.}

\end{center}
\end{table}

\section{Analytical Optimizer Code}
\label{ap_aoc}
Analytical Optimizer is a python code that uses symbolic operations to find the optimum of a quadratic function of any number of variables boxed in linear constraints. It also supports having an unknown free constant in the constraints (not more than one for v1.0) and returns different optimums for different intervals of possible values of the constant. The code supports parallel processing using MPI. 

The main steps that it takes to find the optimum are:
\begin{itemize}
\item In the first step "intersect()" function generates a list of functions which are intersections of the input function with all the possible combinations of constraints. The intersection of each particular combination of constraints is found by solving system of the equations of that combination of the constraints with "solve" function of SymPy package (solution can be underdetermined). Then the solution is plugged into the main function to find the intersection.     

\item In the next step "optimum()" function calculates partial derivatives of each intersection function in the list with respect to all the variables and sets those to zero to find stationary points and their function values (can be a function of the constant in the constraints). This is done with "diff()" and "nonlinsolve()" functions of SymPy package. If the solution is underdetermind, the program discards the solution since this means that the solution is a ridge and it manifests itself as a point at one or more places on the other boundaries.  

\item Next the function "feasible()" checks that those optimum points are feasible (compatible with constraints). With no constants this is a yes or no question but with a constant the answer might be feasible for some values of the constant and not feasible for some other values. This step returns a feasible interval for the constant if we have a constant.

\item In the final step the function "max()" compares all the optimums in their feasible regions to find the maximum for different intervals of the constant (if there is no constant it just returns the optimum function value and the corresponding input values).
\end{itemize}

The code can be downloaded from \cite{github_Moji}.

\end{document}